\documentclass{aa}
\usepackage{graphics}
\usepackage{latexsym}
\usepackage{astron}
\newcommand{\ansteebarklem}{Anstee \& O'Mara (1995) and
Barklem \& O'Mara (1997)}
\newcommand{\barklem}{Barklem \& O'Mara (1997)}

\newcommand{\drakenlte}{Drake (1991)}
\newcommand{\feltzinggustafsson}{Feltzing \& Gustafsson (1998)}
\newcommand{\smith}{Smith (1981)}
\newcommand{\ols}{Olsen (1984)}

\newcommand{\olsii}{Olsen (1994)}
\newcommand{\olsiii}{Olsen (1995)}
\newcommand{\thoren}{Thor\'en (2000)}
\newcommand{\thorenfeltzing}{Thor\'en \& Feltzing (2000)}
\begin{document} 
\title{Removal of the calcium underabundance in cool metal rich \\ 
Galactic disk dwarfs\thanks{Based on observations made at ESO, La Silla}}
\titlerunning{Removal of the calcium underabundance in cool dwarfs}
\author{Patrik Thor\'en \inst{}}
\date{Recieved Mars 9, accepted April 26 2000}
\institute{Uppsala Astronomical Observatory, Box 515, S-751 20, Uppsala, Sweden}
\offprints{Patrik Thor\'en \\ (patrik.thoren@astro.uu.se)}
\thesaurus{1(02.01.3,08.01.1,08.12.1)}
 \maketitle
\begin{abstract}

An apparent Ca underabundance for cool metal rich disk dwarfs was
derived by \feltzinggustafsson\nocite{feltzing}. This was suggested to 
be a NLTE effect, following the prediction by \drakenlte.
New NLTE calculations with MARCS atmospheres and opacities show that
deviations from LTE are very small and can {\it not} explain the
underabundance. It is shown that the underabundance 
was primarily due to erroneously calculated atomic line broadening 
parameters (van der Waals broadening). Part of the underabundance was
also due to the decision not to change photometrically determined 
stellar parameters to satisfy the Fe {\sc i} excitation balance.


\end{abstract}
  

\section{Introduction}

Calcium, being one of the so called $\alpha$-elements, is believed to 
be produced in SNII, exploding massive stars. Due to the early start
of such events in the Galaxy as compared to the other stellar nucleosynthesis 
sites SNIa and AGB stars, old (metal poor) stars are usually rich in Ca
compared to Fe as compared to the corresponding solar Ca/Fe ratio 
(e.g McWillliam et al., 1995; Chen et al., 2000\nocite{mcwilliametal,chen00}).
For metal rich stars the Ca/Fe trend seems to level out at the solar value. 
A considerable scatter was found for 48 metal rich disk dwarfs by
\feltzinggustafsson. As a mean [Ca/Fe]$\sim 0$, but for the cool dwarfs 
a significant underabundance was derived. To examine this, I observed 10 cool disk 
dwarfs with ESO CAT/CES in Nov-Dec 1995. 
The underabundance in Ca for K-dwarfs 
was suspected to be caused by overionization, as predicted 
by \drakenlte\nocite{drake}. A MULTI \cite{multi} NLTE investigation of this with a new Ca 
atom model was therefore performed by 
\thoren\nocite{thoren00cal_nlte}.


\section{Observations and analysis}

The observational data and the LTE analysis are presented in detail in
a separate paper with a wider scope \cite{thoren00cooldwarfs}. 
The stars observed and the 
parameters used in the models are presented in Table \ref{starphottable}.
The LTE line profiles computed by the Uppsala synthetic spectrum code 
SPECTRUM agreed well with the observed spectra. With a new
Ca model atom for the code MULTI it appeared that NLTE effects
for Ca in the cool dwarfs in the sample actually were very small \cite{thoren00cal_nlte}. 
The recent MARCS atmospheres \cite{asplund_marcs} have a higher, more realistic 
amount of UV metal line blocking than the models used by \drakenlte. 
This increases the opacity and decreases the non-local ionizing radiation field. Except
in the line cores no strong NLTE effects could be seen, rather
the Ca/Fe ratio appeared solar, as for the hotter stars in \feltzinggustafsson.
The suggested underabundance, if not real, had to 
have a different origin. 

\begin{table}\caption{\label{starphottable} Observed stars and their photometrically 
determined parameters. The microturbulence parameter was set to 1 km/s for all models.}
\begin{tabular}{l l l l}
Star       & $T_{\rm eff}$	& log g	& [M/H] \\
           & (K)	&       &         \\
HD 12235 & 5971 & 4.18 & 0.15   \\
HD 21197 & 4457 & 4.59 & 0.13   \\
HD 23261 & 5132 & 4.44 & 0.06   \\
HD 30501 & 5174 & 4.54 & 0.13   \\
HD 31392 & 5390 & 4.29 & 0.06   \\
HD 32147 & 4625 & 4.57 & 0.17   \\
HD 42182 & 4917 & 4.54 & 0.25   \\
HD 61606 & 4833 & 4.55 & 0.06   \\
HD 69830 & 5484 & 4.34 & 0.10   \\
HD 213042 & 4560 & 4.58 & 0.06   \\
The Sun  & 5780 & 4.44 & 0.00       
\end{tabular}
\end{table}

\section{Atomic line data}
The lines used in this analysis are presented in Table \ref{linetable}. 
They were selected from the VALD database \cite{vald}.
Most oscillator strengths of the lines used in the analysis were modified,
 to fit our solar observations. VALD provides oscillator
strengths typically correct to the order of magnitude. The van der Waals broadening 
parameters for lines used in the analysis were also changed. 
For Ca lines the van der Waals broadening width in terms of $\delta \Gamma_6$ 
factors (being the corrections to the classical Uns\"old value)
calculated from lab measurements by \smith\nocite{smith_sunproc:81}
are used, except for the 6798 {\AA } line for which no $\delta \Gamma_6$ value
was available. However, this line is very weak and is not 
affected significantly by pressure broadening. Its broadening was
calculated according to \barklem\nocite{barklem:97}. For 
Fe {\sc i} lines the damping was calculated according to 
\ansteebarklem\nocite{anstee,barklem:97} for the lines
in Table 2 marked with asterisks. For the remaining Fe {\sc i} lines
 the $\delta \Gamma_6$ factor had to be used. 

When the atomic line data for the 
project were examined, the explanation was highlighted.
The Uppsala code EQWIDTH had been used for
the analysis in \feltzinggustafsson. This code uses
a correction factor $\delta \Gamma_6$ as input for atomic 
line broadening by H atoms. To get the broadening 
the factor is multiplied with the classical Uns\"old 
broadening value. This factor is typically $\sim 2$. 
The Ca atomic line parameters used by 
\feltzinggustafsson \ and  \thoren\nocite{thoren00cal_nlte} are
both taken from \smith\nocite{smith_sunproc:81}.
The Ca lines $\delta \Gamma_6$ factors were, however, erroneously calculated 
in \feltzinggustafsson, causing the line broadening to be much 
too large. In the cool dwarfs the overall line strength is larger
than in the hotter ones. For lines with equivalent widths increasing beyond 50
m{\AA } this parameter soon becomes very important, because
it brings the line out of the saturated region on the curve of 
growth faster. This means that a model line is too strong for any given
abundance, forcing a reduction of the abundance in order to satisfy
the measured equivalent width. 

Since \feltzinggustafsson \ used equivalent widths, rather than 
synthetic spectra, the effect was not noticed until it 
showed up as an apparent underabundance for Ca in the cool dwarfs. 
Since such a NLTE effect had already been predicted \cite{drake},
this was suggested to be the cause. Because of this it was decided
to make the new K dwarf observations and a new MULTI analysis for Ca.
The Ca NLTE properties of cool metal rich dwarfs will be discussed in
\thoren.

 \begin{figure}                                           
  \resizebox{90mm}{!}{\includegraphics{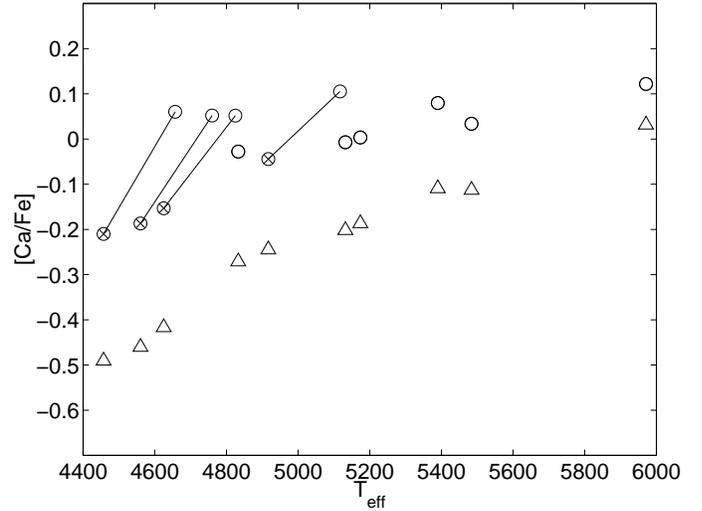}}
  \hfill
  \parbox[b]{90mm}{
    \caption{$\bigtriangleup$ : Ca abundance pattern for 10 dwarfs with old $\delta \Gamma{_6}$ values, 
   $\bigcirc$ : abundances with new $\delta \Gamma{_6}$ values. 
   $\bigotimes$ : abundances for HD21197, HD32147, HD42187 and HD213042 before 
   $T_{\rm eff}$ adjustment. $\bigcirc$ : abundances after $T_{\rm eff}$ adjustment.
   Lines connecting different symbols indicate identical stars.}
    \label{gammma6}}
 \end{figure}

The abundances were obtained in the following way: First synthetic
spectra were fitted with SPECTRUM to the observed spectra. The atmospheric
model parameters used were those determined with photometric Str\"omgren
calibrations for Pop I F-G dwarfs, presented by \ols. The 
Str\"omgren colors were obtained from the catalogues in \olsii \ 
and \olsiii.

The equivalent
widths of the fitted synthetic lines to be used in the abundance analysis were
exported from the fitted spectra. These equivalent widths were
then used as measured ones from the observed spectra and used as input
into the code EQWIDTH (which was used in the analysis of \feltzinggustafsson).
The lines used in the analysis are presented in Table \ref{linetable}.
 
\begin{figure}                                           
  \resizebox{90mm}{!}{\includegraphics{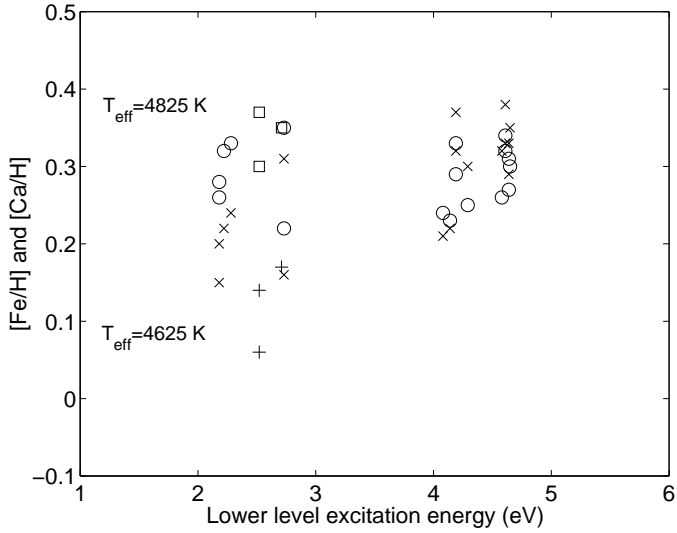}}
  \hfill
  \parbox[b]{90mm}{
    \caption{Abundances for different Fe and Ca lines of HD32147. $\times$ and $+$ : 
    Fe and Ca abundances before temperature 
    correction. $\bigcirc$ and  $\Box$ : Fe and Ca abundances after 
    temperature correction.}
    \label{excbalance}}
 \end{figure}

Figure \ref{gammma6} shows the difference in abundance pattern for the old
and new calculated values of $\Gamma{_6}$. Triangles represent abundances 
determined with the values in \feltzinggustafsson, circles represent abundances
with corrected $\Gamma{_6}$ values. 
As seen in the Fig. \ref{gammma6} there remains an increasing trend
in [Ca/Fe] with increasing temperature after the damping 
treatment correction. This effect is reduced to virtually zero when 
the photometric effective temperatures are adjusted, according to 
\thorenfeltzing, until Fe {\sc i} lines of different excitation 
energy give similar iron abundances. Four of the
stars required such changes. 
Their Ca abundances before ($\bigotimes$) and after ($\bigcirc$) temperature adjustments are shown 
in figure \ref{gammma6}.
The need for $T_{\rm eff}$ changes is illustrated for one of
the objects in Fig. \ref{excbalance}. HD32147 needed a positive
temperature correction of 200 K, which also raised the Ca abundance 
by +0.20 dex.

\section{Summary}

Changing $\Gamma{_6}$ to the 'true' value reduces the underabundance
by ~0.3 dex for the coolest stars in this sample.
This solves the major part of the underabundance problem.

The photometric effective temperatures adopted by \feltzinggustafsson \
 lead in some cases to severe trends in the abundances derived
from Fe {\sc i} lines with different excitation energy. Modifications
of these $T_{\rm eff}$ values dramatically decrease this abundance
scatter and also modify the Ca abundances to values observed
in hotter stars of similar metallicity.

The results in this work and in the forthcoming paper \cite{thoren00cooldwarfs} 
indicate that LTE abundance analysis is indeed useful even for cool metal rich 
dwarfs, increasing the number of objects available for studying
 the chemical evolution of the Galaxy.

\begin{acknowledgement}
The author was supported by the Swedish Natural Science Research Council, NFR.
Professor Bengt Gustafsson is thanked for suggesting the writing of this letter.
Dr. Paul Barklem and Docent Bengt Edvardsson are thanked for valuable comments
on the text.    
\end{acknowledgement}

\begin{table}\caption{\label{linetable} Line data used in this 
investigation. Astrophysical ${\rm log} gf$ values were determined from 
ESO CAT-CES solar flux observations. Asterisks indicates that the pressure 
 broadening was calculated according to \ansteebarklem \ and that
 $\delta \Gamma_6$ was not used.}
\begin{tabular}{l l l l l}
Wavelength  &  $\chi_{\rm low}$  & ${\rm log} gf$ & $\delta \Gamma_6$   &  $\Gamma_{rad}$ \\          
(\AA ) & (eV)               &              &               & (rad s$^{-1}$) \\
Ca I        &           &            &         &       \\                               
  6166.439  &    2.521  &   -1.142   &    1.64 & 1.858E+07     \\     
  6455.598  &    2.523  &   -1.290   &    0.71 & 4.645E+07     \\                 
  6798.467  &    2.709  &   -2.520   &    \ \ * & 1.941E+07     \\               
Fe I 	    &           &            &          &               \\    
  6151.618  &    2.176  &   -3.379   &    \ \ * & 1.549E+08     \\               
  6157.728  &    4.076  &   -1.320   &    \ \ * & 5.023E+07     \\   
  6159.378  &    4.607  &   -1.970   &    \ \ * & 1.919E+08     \\              
  6165.360  &    4.143  &   -1.554   &    \ \ * & 8.770E+07     \\              
  6173.336  &    2.223  &   -2.920   &    \ \ * & 1.671E+08     \\  
  6180.204  &    2.727  &   -2.686   &    1.40 & 1.469E+08     \\   
  6430.846  &    2.176  &   -2.006   &    1.40 & 1.648E+08     \\     
  6436.407  &    4.186  &   -2.410   &    1.40 & 3.041E+07     \\     
  6481.870  &    2.279  &   -2.984   &    1.40 & 1.549E+08     \\            
  6756.563  &    4.294  &   -2.750   &    \ \ * & 7.345E+07    \\     
  6786.860  &    4.191  &   -1.850   &    1.40 & 1.986E+08     \\             
  6804.001  &    4.652  &   -1.546   &    1.40 & 1.758E+08     \\            
  6804.271  &    4.584  &   -1.813   &    1.40 & 5.236E+07     \\        
  6806.845  &    2.727  &   -3.110   &    1.40 & 1.021E+08     \\   
  6810.263  &    4.607  &   -0.986   &    1.40 & 2.301E+08     \\           
  6820.372  &    4.638  &   -1.120   &    1.40 & 2.218E+08     \\            
  6828.591  &    4.638  &   -0.820   &    \ \ * & 2.301E+08                
\end{tabular} 
\end{table}

\end{document}